\documentclass[pre,10pt,twocolumn]{revtex4-1}
\usepackage{amssymb}
\usepackage{latexsym}
\usepackage{amsfonts}
\usepackage{graphics}
\usepackage{graphicx}
\usepackage{epsfig}
\usepackage{amsmath}
\usepackage{float}
\usepackage{verbatim}
\usepackage[subfigure]{graphfig}
\usepackage{color}
\usepackage{soul,xcolor}
\setstcolor{red}
\begin{document}

\title{$k$-core percolation on complex networks: 
\\ Comparing random, localized and targeted attacks}

\author{Xin Yuan,$^1$ Yang Dai,$^2$  H. Eugene Stanley,$^1$ and Shlomo
  Havlin$^{1,3}$} 

\affiliation{$^1$Center for Polymer Studies and Department of Physics,
Boston University, Boston, Massachusetts 02215 USA\\
$^2$School of Economics and Management, Southwest Jiaotong University,
 Chengdu 610031, China\\
$^3$Minerva Center and Department of Physics, Bar-Ilan University,
Ramat-Gan 52900, Israel }

\date{13 May 2016}

\begin{abstract}
The type of malicious attack inflicting on networks greatly influences their 
stability under ordinary percolation in which  a node fails when it becomes 
disconnected from the giant component. Here we study its generalization, 
$k$-core percolation, in which a node fails when it loses connection to a 
threshold $k$ number of neighbors. We study and compare analytically and 
by numerical simulations of $k$-core percolation the stability of networks 
under random attacks (RA), localized attacks (LA) and targeted attacks (TA), 
respectively. By mapping a network under LA or TA  into an equivalent 
network under RA, we find that  in both single and interdependent networks,
TA exerts  the greatest damage to the core structure of a network.  We 
also find  that for Erd\H{o}s-R\'{e}nyi (ER) networks, LA and RA exert equal 
damage to the core structure whereas for scale-free (SF) networks, LA exerts  
much more damage than RA does to the core structure.
\end{abstract}


\maketitle

\section{Introduction}
In complex networks science, malicious attacks may crucially change the 
structure, stability and function of a network
\cite{watts1998collective,albert2000error,cohen2000resilience,callaway2000network,albert2002statistical,newman2003structure,
song2005self,barabasi1999emergence,
caldarelli2007edited,cohen2010complex,rosato2008modelling,arenas2008synchronization,
  newman2010networks,li2010towards,schneider2011mitigation,bashan2012network,dorogovtsev2013evolution,
  ludescher2013improved,yan2013efficient,boccaletti2014structure,li2015percolation,Filippo2015,morone2015influence}. 
The description of an attack on a network is often represented by the 
ordinary percolation model in which the giant connected component 
serves as the relevant order parameter that shows the robustness of 
a macroscopic cluster. The behavior of the giant connected component 
is characteristic of the structural transition of networks where nodes 
suffer either random attacks (RA)
\citep{albert2000error,cohen2000resilience,callaway2000network,buldyrev2010catastrophic,peixoto2012evolution,baxter2012avalanche,
bashan2013extreme,radicchi2013abrupt},
localized attacks (LA) \cite{shao2015percolation, berezin2015spatially,yuan2015breadth}
or targeted attacks (TA) \cite{albert2000error,cohen2000resilience, huang2011robustness,dong2012percolation}.

A natural generalization of ordinary percolation is the $k$-core percolation 
in which the behavior of the $k$-core characterizes the structural 
change of a network under RA \citep{dorogovtsev2006k,goltsev2006k,azimi2014k}. 
The $k$-core of a network is defined as the largest subgraph in which 
each node has at least $k$ neighbors and is obtained through the 
pruning process in which nodes of degree less than $k$ are progressively 
removed.  If $k=1$, then the $k$-core is simply 
the connected component of the network and the giant $k$-core
is the giant connected component, exactly as in ordinary 
percolation. If $k = 2$, we again have a continuous transition similar to 
ordinary percolation, as the 2-core is obtained by simply pruning all dangling 
branches from the 1-core \cite{dorogovtsev2006k,baxter2011heterogeneous}.
Under the $k$-core percolation with $k\geq 3$, single networks 
demonstrate discontinuous transitions at a $k$-dependent critical threshold 
$p_c(k)$ \citep{dorogovtsev2006k,goltsev2006k,azimi2014k}. 
Although prior research has developed tools for probing network
resilience against RA in the context of $k$-core percolation,  and 
has found that degree distribution strongly influences network stability
\cite{dorogovtsev2006k,goltsev2006k}, a systematic study of how
TA and LA affect network resilience in the case of $k$-core percolation
is still missing.

Here we extend the general formalism of the $k$-core percolation for
uncorrelated networks with arbitrary degree distributions under RA 
\cite{dorogovtsev2006k,goltsev2006k} to networks under LA and TA, 
respectively. This allows us to obtain the sizes and other structural 
characteristics of $k$-cores in a variety of damaged random networks 
and to compare the robustness of the networks under these three types 
of attack scenarios in terms of $k$-core percolation. 

We apply our derived general frameworks to study (i) single ER
networks \cite{erdds1959random,bollobas1998random} with a Poisson 
distribution, (ii) single SF networks \cite{barabasi1999emergence,caldarelli2007edited,cohen2010complex}
with a power-law distribution, (iii) two interdependent ER networks with
the same Poisson distribution in each network, and (iv) two interdependent 
SF networks with the same power-law distribution in each network. For 
each case, we investigate how the type of attack influences the $k$-core 
percolation properties. These include the size of the $k$-core, $M_k(p)$, 
as a function of $p$, the fraction of unremoved nodes and the critical 
threshold $p_c(k)$ at which the $k$-core $M_k(p)$ first collapses. 
In all cases we find that our extensive simulations and analytical calculations 
are in good agreement. In general, TA exerts the biggest destruction on 
the $k$-core structure of networks since the hubs of the networks$-$nodes 
with higher degrees$-$are more likely to be removed initially.  We observe 
similar characteristics of robustness in both single and interdependent ER 
networks under both LA and RA. However, for SF networks, LA exerts considerably 
more damage than RA does to the core structure.

\section{RA, LA and TA on a Single Network}

\subsection{Theory}
(\textrm{I}) \textit{Random Attack:}  Following Ref.~\cite{newman2002spread}, 
we introduce the generating function of the degree distribution $P(q)$ of a random 
network $A$ as
\begin{equation}
G_0(x)=\sum_qP(q)x^{q}.
\end{equation}

After an initial attack which is manifested by the random removal of a 
fraction $1-p$ of nodes from the network of size $N$, a cascading pruning 
process occurs as nodes with degree less than $k$ are progressively 
disconnected from the network.  We denote the stage right after the 
random attack as stage $t=0$ and the probability that a given end of 
an edge is the root of an infinite ($k$-1)-ary subtree  as $f_0$ \cite{dorogovtsev2006k}. 
After the first round of pruning process which disconnects those nodes 
with active degree less than $k$ to the rest of network, we obtain a network
in which a fraction $1-p$ of nodes failed due to initial attack and some 
other fraction of nodes  have become isolated due to $k$-core percolation. 
Now this network is at stage $t=1$ and at this time $f_0$ decreases to 
$f_1$. Note an end of an edge is a root of an infinite ($k$-1)-ary subtree 
if at least $k-1$ of its children's branches are also roots of infinite ($k$-1)-ary 
subrees \cite{dorogovtsev2006k}. This leads to the equation for $f_1$ in 
terms of $f_0$, which is
\begin{eqnarray}
\nonumber f_1&=&p\sum_{q=k-1}^{\infty}\frac{P(q+1)(q+1)}{\left\langle q\right\rangle} \sum_{j=k-1}^q C_{q}^{j} f_0^{j}(1-f_0)^{q-j} \\
&\equiv& p\Phi(f_0),
\end{eqnarray}
where $C_{q}^{j}=q!/(q-j)!j!$, $p$  is the probability that the end of the edge is 
occupied, ${P(q+1)(q+1)}/{\left\langle q\right\rangle}$ is the probability that 
a randomly chosen edge leads to a node with $q$ out-going edges (other than 
the one first chosen) and $C_{q}^{j} f_0^{j}(1-f_0)^{q-j}$ is the probability 
that $j$ out of these $q$ branches are roots of infinite ($k$-1)-ary subrees.  
Note that $j$ here must be at least equal to $k-1$.

Similarly, after the pruning process finishes for the second time, we would have 
$f_2=p\Phi(f_1)$. More generally, at each stage $t$ we have $f_t$ obtained 
from $f_{t-1}$ through
\begin{equation}
f_t=p\Phi(f_{t-1}), \label{recursive}
\end{equation}
and the probability that a random node in the damaged network belongs to 
the $k$-core is \cite{dorogovtsev2006k}
\begin{eqnarray}
\nonumber [M_k(p)]_t &=&p\sum_{q=k}^{\infty}P(q)\sum_{j=k}^{q}C_{q}^{j}{f_t}^j (1-f_t)^{q-j} \\
&\equiv& p\Psi(f_t). \label{ra_core}
\end{eqnarray}
 Note that $[M_k(p)]_t$ is also the normalized size of the $k$-core of the 
 network at this stage. As $t\rightarrow \infty$, the network will reach a 
 steady state and we have $f_t \rightarrow f$, with $f$ satisfying the self-consistent 
 equation
\begin{equation}
f=p\Phi(f). \label{equation_f}
\end{equation}
Note an equivalent equation for $f$ at the steady state was also given
in Eq.~(2) of Ref.~\cite{dorogovtsev2006k}.

 We note that for any given $p$, $f$ can be solved from Eq.~(\ref{equation_f}) 
 using  Newton's method with a proper initial value.  A trivial solution $f=0$ exists 
 if the occupation probability $p$ is small and thereafter $M_k(p)=0$,
 i.e., no $k$-core exists in this case. As $p$ increases and at $p=p_c^{RA}(k)$, 
 a non-trivial solution $f=f_c \neq0$ first arises and gives birth to a  $k$-core. 
 This is typical first-order phase transition behavior for the network
 and it requires the derivatives of both sides of Eq.~(\ref{equation_f}) with respect 
 to $f_c$ be  equal \cite{dorogovtsev2006k,goltsev2006k}, i.e.,
 \begin{equation}
 1=p_c^{RA}(k)\Phi^{'}(f_c).  \label{equation_fc}
 \end{equation}
Therefore by using Eqs.~(\ref{equation_f}) and (\ref{equation_fc}), the 
threshold of $k$-core percolation $p_c^{RA}(k)$ is determined by
\begin{align}
p_c^{RA}(k)=1/\Phi^{'}(f_c),   \quad f_c=\Phi(f_c)/\Phi^{'}(f_c).
\end{align}
Here, $f_c$ is the value of $f$ at the birth of a $k$-core. When
$p>p_c^{RA}(k)$, there is always a non-zero solution of $f$ that ensures 
the existence of a  $k$-core.

(\textrm{II}) \textit{Localized Attack:} 
We next consider the localized
attack on network $A$ by the removal of a fraction $1-p$ of nodes, starting 
with a randomly-chosen seed node. Here we remove the seed node and its 
nearest neighbors, next-nearest neighbors, next-next-nearest neighbors, and 
continue until a fraction $1-p$ of nodes have been removed from the network. 
This pattern of attack reflects such real-world localized scenarios as earthquakes 
or the results of weapons of mass destruction.  As in Ref.~\citep{shao2015percolation}, 
the localized attack occurs in two stages, (i) nodes belonging to the attacked 
area (the seed node and the layers surrounding it) are removed but the links 
connecting them to the remaining nodes of the network are left in place, but 
then (ii) these links are also removed. Following the method introduced in
Refs.~\cite{shao2015percolation,shao2009structure}, we find the
generating function for the degree distribution of the remaining network to be
\begin{equation}
G_{p0}(x)=\frac{1}{G_0(l)}G_0[l+\frac{G^{'}_0(l)}{G^{'}_0(1)}(x-1)],\label{Gp0}
\end{equation}
where $l\equiv G^{-1}_0(p)$.  

Next we want to find an equivalent network $\tilde{A}$ 
such that a random removal of a fraction $1-p$ of nodes from it will produce
a network with the same degree distribution as that obtained by a LA on 
network $A$ described above.  We denote $P(q^{'})$ as the degree distribution 
of network $\tilde{A}$ and $\tilde{G}_{A0}(x)$ as its generating function.
Following the argument of equivalence discussed above and
by setting $\tilde{G}_{A0}(1-p+px)=G_{0}^{p}(x)$ \cite{newman2010networks,yuan2015breadth},
and after some rearrangement, we  have $\tilde{G}_{A0}(x)$  as
\begin{equation}
\tilde{G}_{A0}(x)=\frac{1}{G_0(l)}G_0[l+\frac{G_0^{'}(l)}{G_0^{'}(1)G_0(l)}(x-1)]. \label{G_A_tilde}
\end{equation}
Therefore, $P(q{'})$ could be generated from $\tilde{G}_{A0}(x)$ through 
direct differentiation \cite{newman2010networks}
\begin{equation}
P(q{'})=\frac{1}{q{'}!}\frac{d^{q{'}}}{dx^{q{'}}}\tilde{G}_{A0}(x). \label{derivative}
\end{equation}
 Combining Eqs.~(\ref{G_A_tilde}) and (\ref{derivative}) we obtain 
the degree distribution of the equivalent network $\tilde{A}$ as
\begin{equation}
P(q{'})=\sum_{q=q{'}}^{\infty}\frac{l^q}{p}P(q)C_{q}^{q{'}}(\frac{\tilde{p}}{p})^{q{'}}(1-\frac{\tilde{p}}{p})^{q-q{'}}, \label{P(a)_LA}
\end{equation}
with $\tilde{p}=G_0^{'}(l)/G_0^{'}(1) l$.

Thus performing $k$-core percolation on the resultant network after LA
is equivalent to performing $k$-core percolation on network $\tilde{A}$
after a random removal of the same fraction of nodes.  This enables 
us to transform a LA problem into the familiar RA problem examined in 
the previous scenario. Then for the LA scenario we replace $P(q)$ in 
Eqs.~(\ref{ra_core}) and (\ref{equation_f}) with $P(q{'})$ obtained 
from Eq.~(\ref{P(a)_LA}) and obtain the size of $k$-core $M_k(p)$ 
as well as its critical threshold $p_c^{LA}(k)$.

(\textrm{II}I) \textit{Targeted Attack:} 
Next, we consider the targeted attack on network $A$ by the
removal of a fraction $1-p$ of nodes where nodes are removed 
based on their degree \cite{huang2011robustness,dong2012percolation}. 
This pattern of attack reflects such real-world cases as intentional attacks 
on important transportation hubs or sabotage on the Internet \cite{cohen2001breakdown}.  
To analyze this case, a value $W_{\alpha}(q_i)$ is assigned to each node, 
which represents the probability that a node $i$ with $q_i$ links is initially 
attacked and becomes dysfunctional. This probability is described through 
the family of functions \cite{gallos2005stability}
\begin{equation}
W_{\alpha}(q_i)=\frac{q_i^{\alpha}}{\sum_{i=1}^{N}q_i^{\alpha}}, -\infty <\alpha <+\infty. \label{Prob_TA}
\end{equation}
When $\alpha >0$, nodes with higher connectivity have a higher probability 
to be removed while $\alpha <0$ indicates otherwise.  Note that for $\alpha=0$, 
all nodes have equal probability to be removed, which is exactly the
same as the RA case. 

As described in Ref.~\cite{huang2011robustness}, the targeted attack occurs
in two stages, (i) nodes are chosen according to Eq.~(\ref{Prob_TA}) and later 
removed but the links connecting the removed nodes and the remaining nodes 
are left in place, but then (ii) these links are also removed. 

Following the method introduced in Refs.~\cite{huang2011robustness, shao2009structure}, 
we find the generating function for the degree distribution of the remaining 
network to be (only removing the nodes)  
\begin{equation}
G_{b}(x)=\frac{1}{p}\sum_q P(q)l^{q^{\alpha}} x^q, \label{G_til_b}
\end{equation}
where $l=G_{\alpha}^{-1}(p)$ and $G_{\alpha}(x)\equiv \sum_{q=0}^{\infty}P(q)x^{q^{\alpha}}$.
 The fraction of the original links that connect to the remaining nodes is 
 $\tilde{p}={\sum_q P(q)q l^{q^{\alpha}}}/{\sum_q P(q) q}$. 
Further removing the links  which end at the removed nodes of a randomly 
connected network is equivalent to randomly removing a fraction $1- \tilde{p}$ 
of links of the remaining nodes. Using the approach introduced in 
Ref.~\cite{newman2010networks}, we find that the generating function of 
the remaining nodes after the removal of the links between removed 
nodes and remaining nodes is
\begin{equation}
G_{c}(x)=G_{b}(1-\tilde{p}+\tilde{p}x).  \label{G_c}
\end{equation}
Next we find an equivalent network $\tilde{B}$ in which a random 
removal of a fraction $1-p$ of nodes will produce a network with the 
same degree distribution as that obtained by a TA on network $A$ 
described above.  We denote $P(q^{'})$ as the degree distribution 
of network $\tilde{B}$ and $\tilde{G}_{B0}(x)$ as its generating 
function. Following the equivalence argument discussed above and 
setting $\tilde{G}_{B0}(1-p+px)=G_{c}(x)$ \cite{newman2010networks},
after some algebra, we obtain $\tilde{G}_{B0}(x)$ as $\tilde{G}_{B0}(x)=G_{c}(1+\frac{1}{p}(x-1))$. 
Using Eq.~(\ref{G_c}), we thus have 
\begin{equation}
\tilde{G}_{B0}(x)=G_{b}(\frac{\tilde{p}}{p}(x-1)+1). \label{G_B_tilde}
\end{equation}
Accordingly, combining Eqs.~(\ref{G_til_b}) and (\ref{G_B_tilde}) and 
using direct differentiation we obtain the degree distribution $P(q{'})$
of the equivalent network $\tilde{B}$ as 
\begin{equation}
P(q{'})=\sum_{q=q{'}} ^{\infty}\frac{l^{q^{\alpha}}}{p}P(q)C_{q}^{q{'}}(\frac{\tilde{p}}{p})^{q{'}}(1-\frac{\tilde{p}}{p})^{q-q{'}}. \label{P(a)_TA}
\end{equation}

Thus performing $k$-core percolation on network $A$ after a TA is 
the same as performing the $k$-core percolation on network $\tilde{B}$ 
after a random removal of the same fraction of nodes.  By replacing $P(q)$ in 
Eqs.~(\ref{ra_core}) and (\ref{equation_f}) with $P(q{'})$ obtained from 
Eq.~(\ref{P(a)_TA}), for the TA scenario we can obtain the size of $k$-core
$M_k(p)$ together with its critical threshold $p_c^{TA}(k)$.

\subsection{Results}
To test the analytical solutions derived in Section A, we conduct numerical
solutions of the analytic expressions, and compare the results with
simulation results on single networks with degrees following both
Poisson distributions and power-law distributions under RA, LA and
TA. All the simulation results are obtained for networks with $N=10^6$
nodes.

\subsubsection{ Erd\H{o}s-R\'{e}nyi networks}
We first consider ER networks of which the degree distribution is Poissonian, 
i.e., $P(q)=e^{-\lambda}\frac{\lambda^q}{q!}$ with the average degree denoted 
by $\lambda$.
\begin{figure}
\includegraphics[width=0.45\textwidth, angle = 0]{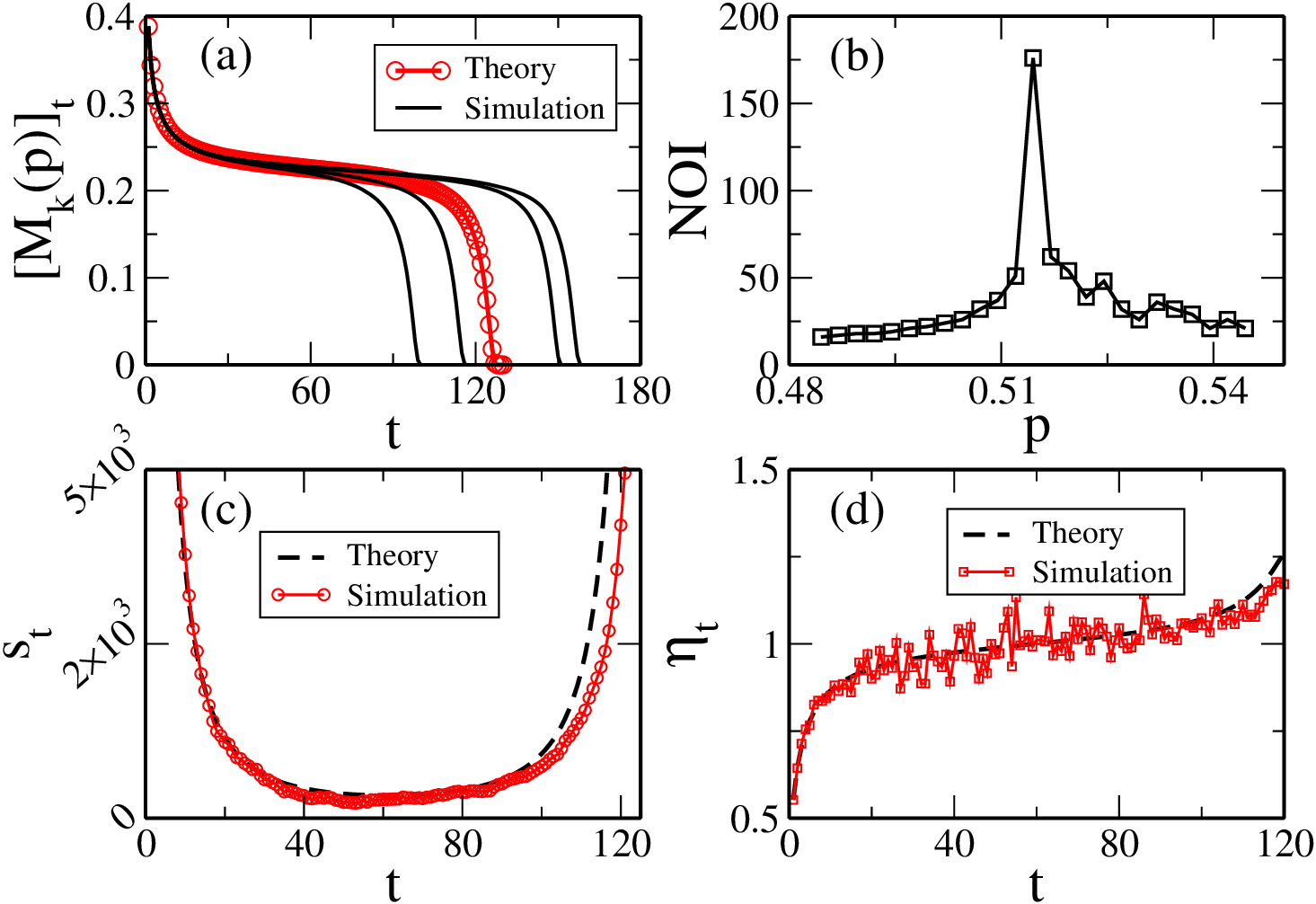}
\caption{\label{Fig1}(Color online) Dynamics of an ER network near 
criticality under random attack applying $k$-core percolation. (a) 
Dynamical process of the $k$-core size $[M_k(p)]_t$ of the ER network 
with $k=4$, $\lambda=10$ and $N=10^6$ both in theory (red line 
with circles) and in simulation (solid black lines) at $p=0.5145$, slightly 
below $p_c^{RA}(k)=0.515$. (b) Number of iterations (NOI) before 
network reaching stability. This number peaks at $p=p_c^{RA}$ and it 
drops quickly as $p$ moves away from $p_c^{RA}(k)$ \cite{parshani2011critical,zhou2014simultaneous}.
(c) At $p=0.5145$, the red line with circles represents the variation 
of failure sizes $s_t$ (only the plateau stage) for one realization in 
the simulation;  the black dashed  line shows $s_t$ for the theoretical 
case. (d) At $p=0.5145$, the red line with rectangles shows the variation 
of the average branching factor $\eta_t$ for one realization in the simulation;
the black dashed line shows $\eta_t$ of the analytic solution. 
Note that this figure is similar to that found in interdependent networks \cite{zhou2014simultaneous}.}
\end{figure}

In the RA scenario on an ER network with $k=4$ and $\lambda=10$, 
we exhibit in Fig.~\ref{Fig1}(a) several realizations the cascading pruning 
process under $k$-core percolation with $p$ slightly smaller than $p_c^{RA}(k)$, 
in comparison with theory.  Note that the simulation results for the cascading 
pruning agree well with analytical results from Eqs.~(\ref{recursive}) and (\ref{ra_core}). 
Different realizations give different results due to random fluctuations of the dynamic processes showing 
deviations from the mean field, rendering small fluctuations around the mean-field 
analytical result. To calculate the first-order phase transition point $p_c^{RA}(k)$ 
with good precision, as shown in Fig.~\ref{Fig1}(b), we identify the characteristic 
behavior of the number of iterations (NOI) in the cascading process \cite{parshani2011critical}. 
This gives us $p_c^{RA}(k)=0.515$, corresponding to the peak of the NOI. Figures~\ref{Fig1}(c) 
and \ref{Fig1}(d) show the variation of the pruning size $s_t$, which
is the number of nodes that are pruned at stage $t$, and the branching factor 
$\eta_t$ ($\eta_t=s_{t+1}/s_t$), respectively, in one typical realization that finally 
reached total collapse. Note that $s_t$ initially drops as
 the network is still well connected and thus less nodes are pruned 
per pruning step ($s_{t}>s_{t+1}$). Then the network becomes weak enough and $s_t$ remains at low  and almost constant 
value during the plateau stage while the network 
keeps getting weaker. Finally $s_t$ rises as a failure in the current step leads to 
more than one failure in the next step and results in the total collapse of the network (see Fig.~\ref{Fig1}(c)). 
 Although $s_t$ first decreases,
the ratio of two consecutive pruning sizes, $\eta_t$, increases. 
Specifically $\eta_t$ increases during the initial cascades from below 
1 to  approximately 1 (with some fluctuations) at the plateau, which starts at time $T$ 
when each of the $s_T$ pruned nodes leads, 
on average, to failure of another single node.  This is a stable state, leading to the 
 divergence of $t$ for $N\rightarrow \infty$, where the cascading 
 trees become critical branching processes \cite{zhou2014simultaneous, baxter2015critical}
 with the average time at criticality scales as 
 $N^{1/3}$ \cite{zhou2014simultaneous}. In a finite network of size $N$, however,
 the accumulated failures weaken the network
 step by step and thus $s_t$ starts to rise, leading to the collapse of the system. 
 During this period, $\eta_t$ rises to above 1 as shown in Fig.~\ref{Fig1}(d).

\begin{figure}
\includegraphics[width=0.45\textwidth, angle = 0]{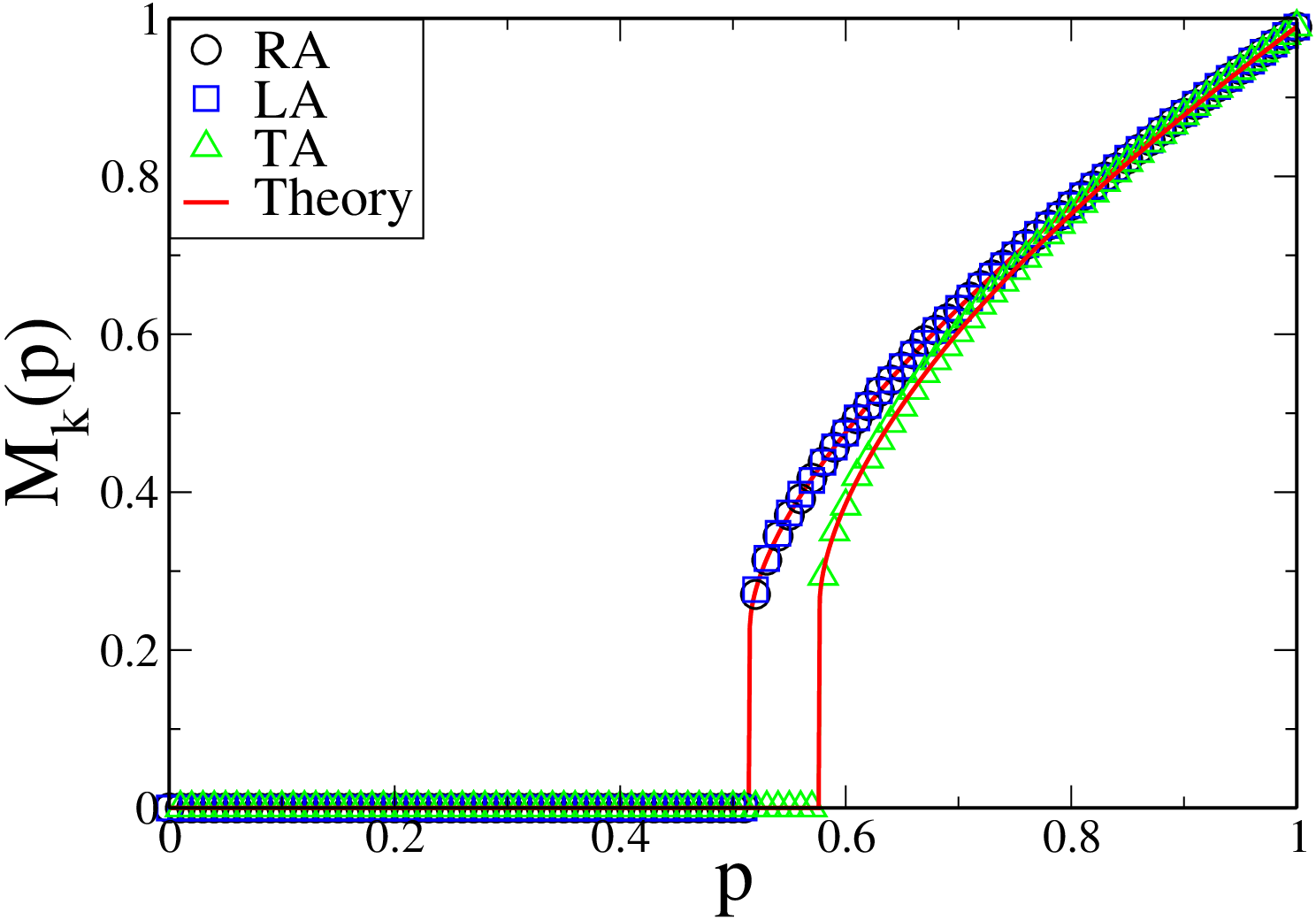}
\caption{\label{Fig2}(Color online) Sizes of the $k$-core,
  $M_k(p)$, as a function of the  fraction of unremoved nodes, $p$, 
  for a single ER network with $\lambda=10$  and $k=4$. Here solid 
  lines are theoretical predictions, from Eq.~(\ref{ra_core}) for RA and
  its counterparts of LA and TA (with $\alpha=1$), and symbols are 
  simulation results with network size $N=10^6$, under RA ($\bigcirc$), 
  LA ($\square$) and TA ($\vartriangle$). Note that for 
  ordinary percolation under either RA or LA, the system is significantly 
  more resilient, and the transition is continuous at $p_c=1/\lambda=0.1$. }
\end{figure}

When the dynamics end, the network enters the steady state. At this state, 
Fig.~\ref{Fig2} shows the $k$-core $M_k(p)$ as a function of the occupation 
probability $p$ under RA, LA and TA (with $\alpha=1$)  in the context of
$k$-core percolation. Note that the simulation results agree well with the 
theoretical results and that there is first-order percolation transition behavior 
in all attack scenarios. Note also that $p^{RA}_c(k)$ is equal to $p^{LA}_c(k)$ 
and they both are smaller than $p^{TA}_c(k)$.  This is similar to ordinary percolation
\cite{huang2011robustness,shao2015percolation}.
This is the case because for ER networks with $P(q)=e^{-\lambda}\frac{\lambda^q}{q!}$,  
from Eq.~(\ref{P(a)_LA}) the degree distribution $P(q{'})$ of the 
equivalent network $\tilde{A}$ can be calculated to be
\begin{eqnarray}
\nonumber P(q{'})&=&\sum_{q=q{'}}^{\infty}\frac{l^q}{p}P(q)C_{q}^{q{'}}(\frac{\tilde{p}}{p})^{q{'}}(1-\frac{\tilde{p}}{p})^{q-q{'}}\\
\nonumber &=& \frac{e^{-\lambda}[\lambda l \frac{\tilde{p}}{p}]^{q{'}}}{pq{'}!}\sum_{q=q{'}}^{\infty}\frac{[\lambda l (1-\frac{\tilde{p}}{p})]^{q-q{'}}}{(q-q{'})!}\\
\nonumber &=&\frac{e^{-\lambda}[\lambda l \frac{\tilde{p}}{p}]^{q{'}}}{pq{'}!} e^{\lambda l (1-\frac{\tilde{p}}{p})}\\
 &=& e^{-\lambda}\frac{\lambda^{q{'}}}{q{'}!},  \label{p_LA_ER}
\end{eqnarray}
where we use $l=\frac{ln (p)}{\lambda}+1$ and $\tilde{p}=p/l$ for 
simplification.  Note that from Eq.~(\ref{p_LA_ER}) the degree distribution 
of network $\tilde{A}$ is also Poissonian and has the same average degree 
$\lambda$ as the original network.  Thus, we have $p^{RA}_c(k)=p^{LA}_c(k)$ 
as observed.  Similarly from Eq.~(\ref{P(a)_TA}) with $\alpha=1$, we find 
the degree distribution $P(q{'})$ of the equivalent network $\tilde{B}$ to be
\begin{equation}
P(q{'})=e^{-\lambda l^2}\frac{(\lambda l^2)^{q{'}}}{q{'}!} \label{p_TA_ER},
\end{equation}
with $l=\frac{ln (p)}{\lambda}+1$.  Note that from Eq.~(\ref{p_TA_ER})
 the degree distribution of network $\tilde{B}$ is also Poissonian but has a 
smaller average degree $\lambda l^2$ as $l$ is always smaller than 1 
\cite{huang2011robustness}. Compared to that under RA, the removal of 
the same fraction of nodes under TA reduces a larger amount of connectivity 
in the network and therefore, in the context of $k$-core percolation, 
the critical threshold $p^{TA}_c(k)$ is significantly larger than $p^{RA}_c(k)$.

\begin{figure}
 \includegraphics[width=0.45\textwidth, angle = 0]{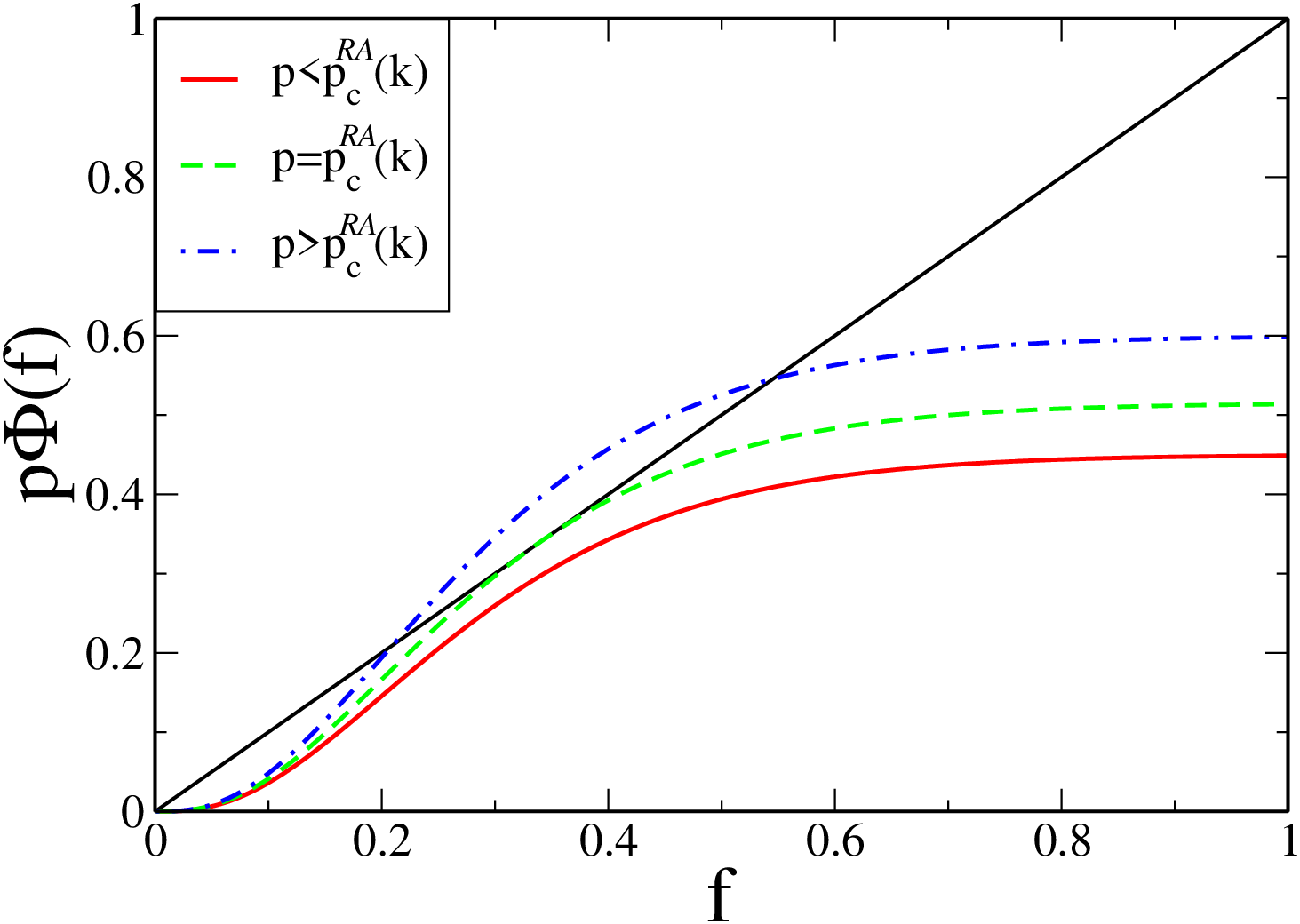}
\caption{\label{Fig3}(Color online) Graphical solution of Eq.~(\ref{equation_f}) 
for the $k$-core percolation with $k=4$ in an ER network under RA with an 
average degree of 10. The straight line and the curves $p\Phi(f)$ show, 
respectively, the left- and right-hand side of Eq.~(\ref{equation_f}) as 
functions of $f$ for different values of $p$. The nonzero solution 
of $f$ appears above the critical value $p_c^{RA}(k)=0.515$, at which 
the right-hand side curve $p\Phi(f)$ starts to intersect the straight line. 
The physical solution is provided by the largest root of the equation $f=p\Phi(f)$ 
when $p>p_c^{RA}(k)$ (the upper intersection in the plot).  }
\end{figure}
As an example, Fig.~\ref{Fig3} shows the solution of Eq.~(\ref{equation_f}) 
for different values of the occupation probability $p$ under RA and 
demonstrates the origin of the first-order transition. When $p<p^{RA}_c(k)$, 
the straight line and the curve only have an intersection at $f=0$, which 
always renders $M_k(p)=0$ according to Eq.~(\ref{ra_core}). 
A  $k$-core $M_k(p)$ first arises discontinuously at $p=p^{RA}_c(k)$, 
when the straight line and the curve tangentially touch each other at 
a nonzero intersection at $f= f_c$, satisfying Eq.~(\ref{equation_fc}).  
As $p$ increases further and becomes greater than $p^{RA}_c(k)$,  $M_k(p)$ 
continues to exist as an additional intersection appears, and this serves as the 
physical solution of $f$ (see the upper intersection in Fig.~\ref{Fig3}). Similar 
procedures are applied to the LA and TA scenarios as well and the corresponding 
$p^{LA}_c(k)$ and $p^{TA}_c(k)$ are obtained, respectively. 
\begin{figure}
\includegraphics[width=0.45\textwidth, angle = 0]{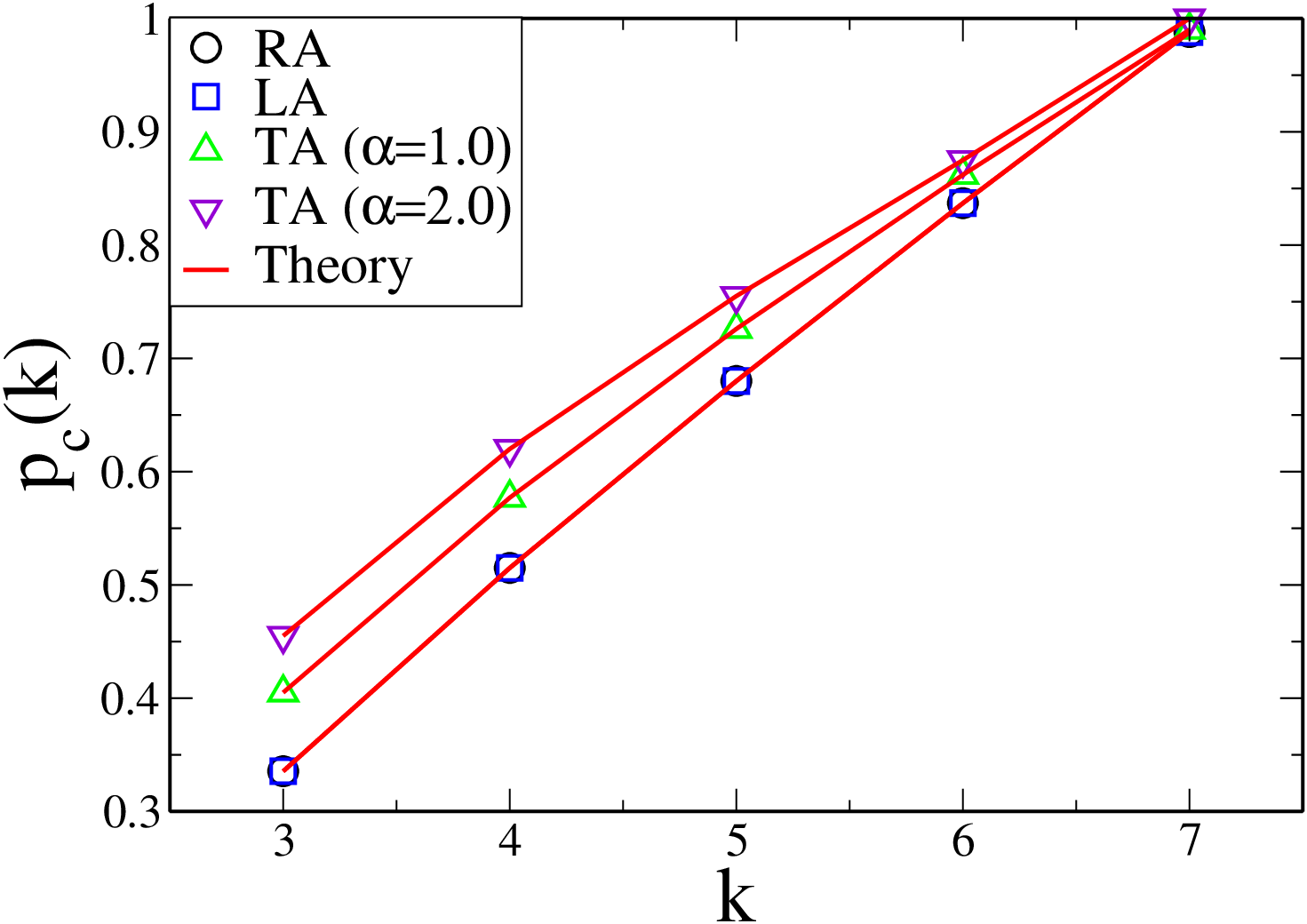}
\caption{(Color online)\label{Fig4} Percolation thresholds $p_c(k)$ of 
a single ER network as a function of $k$ under RA, LA and TA with $\alpha =1$, 
 $\lambda=10$.  Here solid lines are theoretical predictions and symbols ($\bigcirc$ 
 for RA,  $\square$ for LA, $\bigtriangleup$ for TA with $\alpha=1.0$ and
$\bigtriangledown$ are for TA with $\alpha=2.0$) are simulation results with 
 network size of $N=10^6$ nodes. Note that LA coincides with RA.}
\end{figure}

Next we obtain the relationship between the robustness of the network under the
three types of attacks and the threshold $k$ in the context of $k$-core percolation.   
Figure~\ref{Fig4} shows how the percolation thresholds $p_c(k)$ 
under RA, LA and TA, change with $k$ where $\lambda=10$ 
for a single ER network.  Here in Fig.~\ref{Fig4}, as $k$ 
increases from 3  to 7,  $p^{RA}_c(k)$, $p^{LA}_c(k)$ and $p^{TA}_c(k)$
increase accordingly. For each $k$ value,  $p^{RA}_c(k)=p^{LA}_c(k)<p^{TA}_c(k,\alpha=1.0)<p^{TA}_c(k,\alpha=2.0)$,
which indicates that in the context of $k$-core percolation RA and LA cause the same amount of
damage to the structure of an ER network, but that TA causes more severe structural damage 
to an ER network. Moreover, we find that RA and LA have very similar dynamic properties 
in terms of NOI as well as the pruning size $s_t$. Figure~\ref{Fig4} also indicates that with a larger $\alpha$, TA will 
cause more damage since higher degree nodes are more likely to be removed. Similar 
results are reported in the context of ordinary percolation 
on ER networks \cite{huang2011robustness,shao2015percolation}. 

\subsubsection{Single scale-free networks}
\begin{figure}
\includegraphics[width=0.45\textwidth, angle = 0]{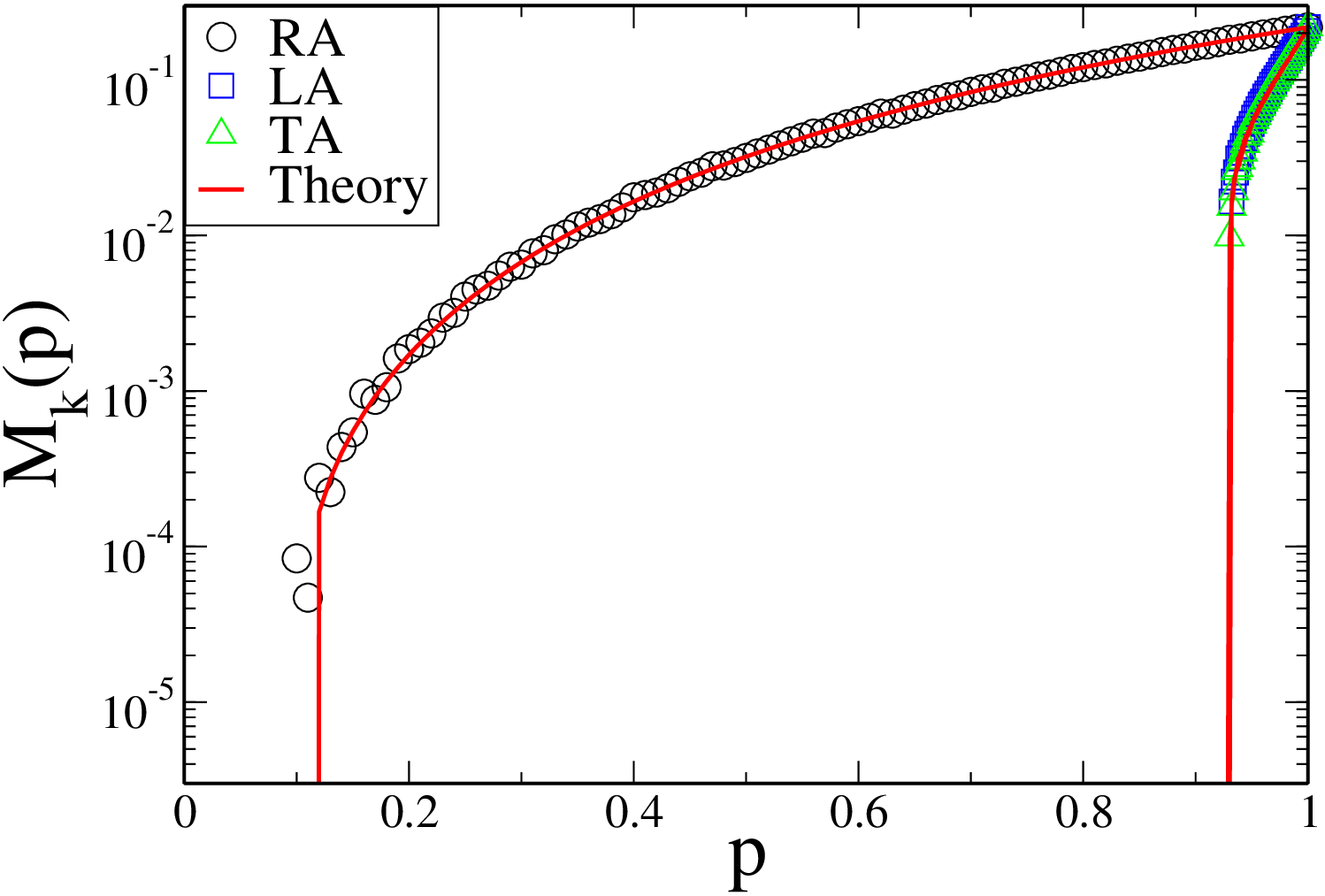}
\caption{(Color online)\label{Fig5} Sizes of the $k$-core,  $M_k(p)$, as a function 
 of the fraction of unremoved nodes, $p$, for a single SF network with $\gamma=2.3$, 
 $q_{min}=2$, $q_{cut}(N)=1000$ and $k=4$. Here solid lines are theoretical predictions, 
 from Eq.~(\ref{ra_core}) for RA and its counterparts of LA and TA (with $\alpha=1$), 
 and symbols are simulation results with network size $N=10^6$, under RA ($\bigcirc$), 
 LA ($\square$)  and TA ($\bigtriangleup$).}
\end{figure}

 We next consider SF networks in which
 degrees of nodes follow a power law distribution,
 i.e., $P(q) \propto q^{-\gamma}$ with the degree exponent $\gamma \in (2,3]$.
As in Ref.~\cite{dorogovtsev2006k},
a size dependent cutoff  $q_{cut}(N)$ of the degree distribution is introduced.
For the configuration model without multiple connections the dependence
$q_{cut}(N) \sim \sqrt{N}$ is usually used when $2<\gamma \leq 3$, and 
first-order percolation transition behavior was observed in the RA case \cite{dorogovtsev2006k}. 
Figure~\ref{Fig5} shows $M_k(p)$ as a function of the occupation 
probability $p$ under RA, LA and TA (with $\alpha=1$) under $k$-core percolation with 
$k=4$ and $\gamma=2.3$. The simulation results agree well with the theoretical results,
and there is first-order percolation transition behavior in all attack scenarios. Note that 
$p^{LA}_c(k)$ is  approximately equal to $p^{TA}_c(k)$, and that they both are significantly 
larger than $p^{RA}_c(k)$.   Because SF networks are ultrasmall \cite{cohen2003scale, cohen2010complex}, 
the LA process can easily spread from the seed node to high degree hubs in several steps 
and therefore severely disrupts the core structure of the network, 
an outcome similar to that of
the TA process.  This is in marked contrast to the case of ER networks in which the 
majority of nodes have degrees around the average degree and therefore for 
the RA and LA processes, nodes of high degrees are less likely to be reached than 
those in the TA process.

\begin{figure}
\includegraphics[width=0.45\textwidth,angle=0]{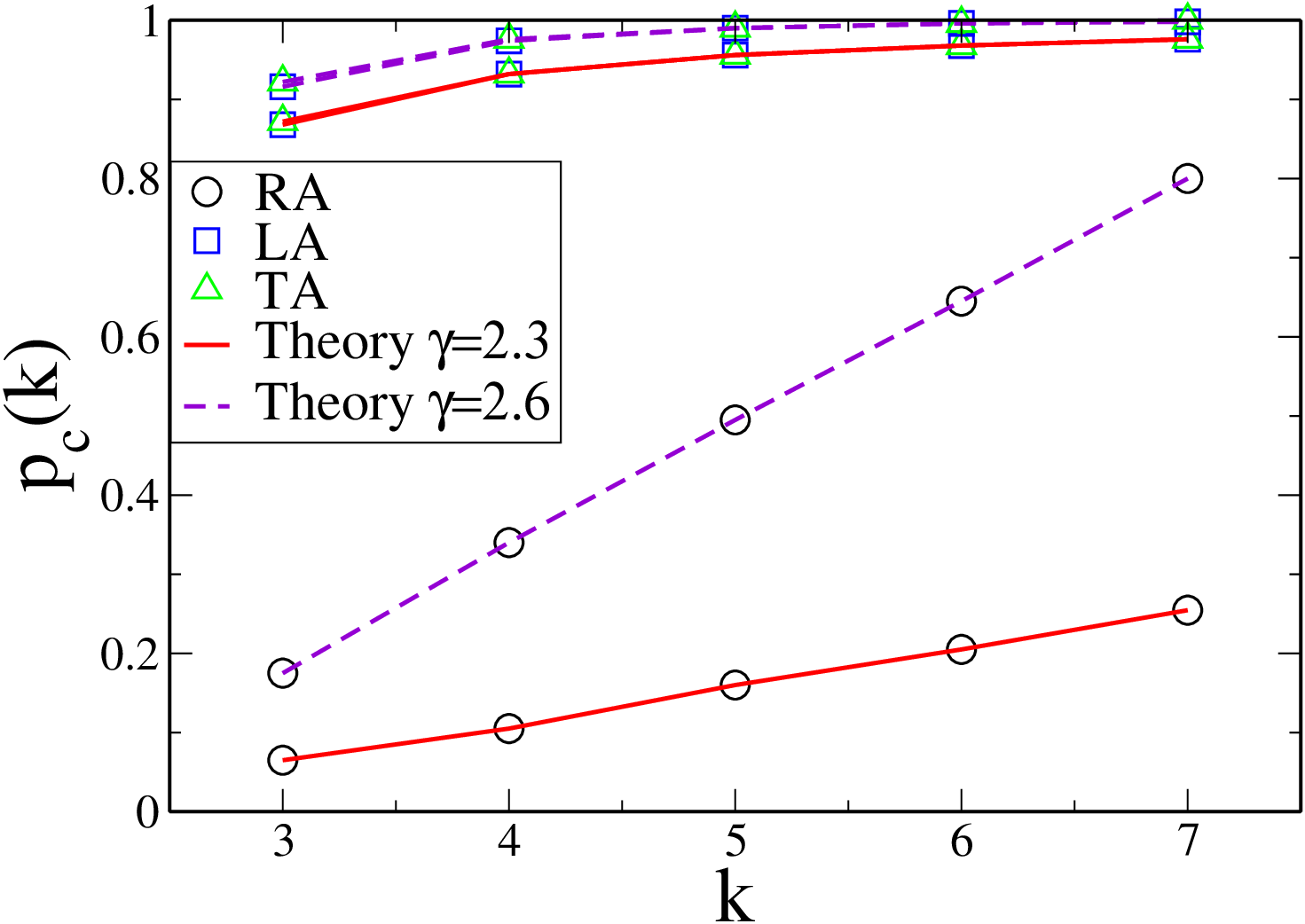}
\caption{\label{Fig6}(Color online) Percolation thresholds $p_c(k)$ of a
  single SF network as a function of $k$ under RA, LA and TA
  with $\alpha =1$, $q_{min}=2$, $q_{cut}(N)=1000$ for $\gamma=2.3$ (solid red lines) 
  and $\gamma=2.6$ (dashed purple lines). Here lines are theoretical 
  predictions and symbols ($\bigcirc$ for RA, $\square$ for LA and $\bigtriangleup$ 
  are for TA) are simulation results with network size of $N=10^6$ nodes. }
\end{figure}

Next we determine the relationship between the robustness of the network under 
three types of attacks and the threshold $k$ in the context of $k$-core percolation.   
For a single SF network, Fig.~\ref{Fig6} shows how the percolation thresholds $p_c(k)$ 
under RA, LA and TA (with $\alpha =1$) change with $k$ for two values of $\gamma$. 
As seen in Fig.~\ref{Fig6}, the $p_c(k)$ values under all attack scenarios for $\gamma=2.3$ 
are smaller than those for $\gamma=2.6$, which indicates that SF networks with smaller 
$\gamma$ values are more stable in the context of $k$-core percolation.  In addition, for each 
value of $\gamma$ as $k$ increases from 3 to 7,  $p^{RA}_c(k)$, $p^{LA}_c(k)$ and 
$p^{TA}_c(k)$ increase accordingly. For each $k$ value, $p^{LA}_c(k)\approx p^{TA}_c(k)>p^{RA}_c(k)$,
which indicates that in the context of $k$-core percolation, LA and TA (with $\alpha=1$)
exert approximately the same amount of damage  to the structure of a SF network whereas 
 RA produces less severe structural damage to a SF network. Analogous results are 
reported in the context of ordinary percolation on SF networks \cite{huang2011robustness,shao2015percolation}. 

\section{RA, LA and TA on Interdependent Networks}

\subsection{Theory}

We extend the formalism of ordinary percolation on fully interdependent 
networks introduced in Ref.~\cite{buldyrev2010catastrophic} to 
$k$-core percolation. Specifically, we consider two networks $A$ and $B$ 
with the same number of nodes $N$. Within each
network the nodes are randomly connected with the same degree distribution
$P(q)$. A fraction $d_A$ of nodes from network $A$ depend on
 nodes in network $B$, and a fraction $d_B$ of nodes from network $B$ depend on 
 nodes in network $A$.  We also assume that if a
node $i$ in network $A$ depends on a node $j$ in network $B$ and node
$j$ depends on node $l$ in network $A$, then $l=i$, which rules out the
feedback condition \cite{gao2013percolation}. This interdependence
means that if node $i$ in network $A$ fails, its dependent node $j$ in 
network $B$ will also fail, and vice versa.

(I) \textit{Random Attack}: We begin by randomly removing a fraction
$1-p$ of nodes in network $A$. All the nodes in network
$B$ that are dependent on the removed nodes in network $A$ are also
removed.  Then a cascading pruning process begins, 
and nodes with degree less than $k_1$ in network $A$ and 
$k_2$ in network $B$ are sequentially removed in the $k$-core percolation 
process. Due to interdependence, the removal process iterates back
and forth between the two networks until they fragment completely or
produce a mutually connected $\textbf{k}$-core with no further
disintegration, where $\textbf{k} \equiv (k_1,k_2)$ \cite{buldyrev2010catastrophic,azimi2014k}.  
\begin{figure}
\includegraphics[width=0.45\textwidth, angle = 0] {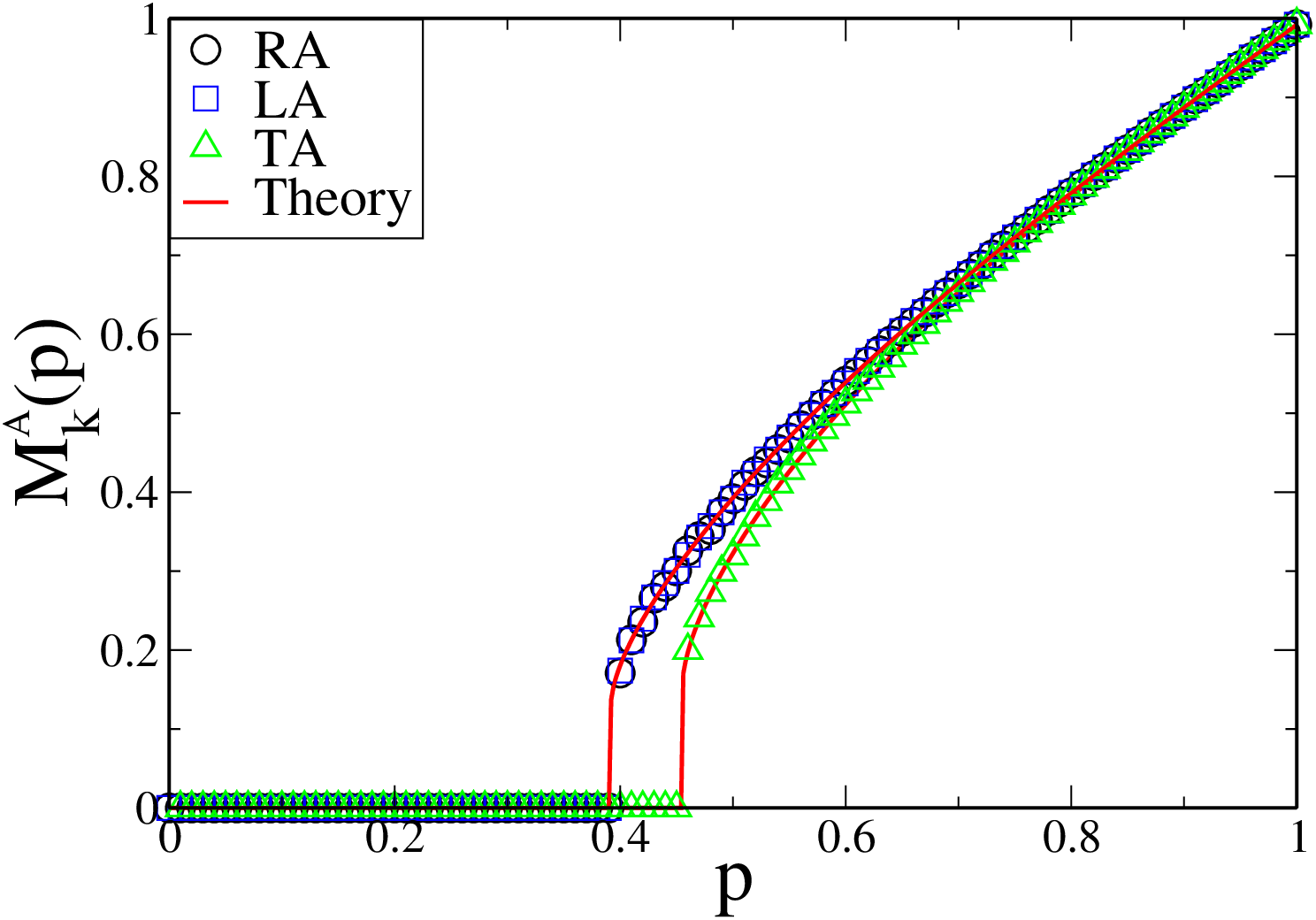}
\caption{\label{Fig7}(Color online) 
 Sizes of $k$-core of network $A$, $M_{\textbf{k}}^A(p)$, as a 
 function of the fraction of unremoved nodes, $p$, for two
  partially interdependent ER networks with $d=0.5$, $\lambda=10$  
  and $\textbf{k}=(3,4)$. Here solid red lines are theoretical predictions, 
  from Eq.~(\ref{giant_cores}) for RA and its counterparts of LA and TA for 
  $\alpha=1$, and symbols are simulation results with network size $N=10^6$, 
  under RA ($\bigcirc$), LA ($\square$) and TA ($\bigtriangleup$).}
\end{figure}

When the system of interdependent networks stops disintegrating, 
as in a single network we let $f_A (f_B)$ be the probability that a given 
end of an edge of network $A (B)$ is the root of an infinite ($k_{1(2)}$-1)-ary subtree.
An end of an edge is a root of an infinite ($k_1$-1)-ary subtree of network $A$ if
it is an autonomous node \cite{parshani2010interdependent} and
at least $k_1-1$ of its children's branches are also roots of infinite ($k_1$-1)-ary 
subrees; otherwise, despite that, the node it depends on has to be in the
$k_2$-core of network $B$. Similar arguments exist for edges in network $B$.  
These lead to the equation of $f_A$ in terms of $f_A$ and $f_B$ as
\begin{eqnarray}\label{f_A_B_1}
\nonumber f_A&=&p\Phi_A(f_A) (1-d_A)+p\Phi_A(f_A)\Psi_B(f_B)d_A \\
&=&p\Phi_A(f_A)\left[ (1-d_A)+d_A\Psi_B(f_B)\right],
\end{eqnarray}
where $p$ is the probability that an end $n_0$ of an edge is occupied,
$\Phi_A(f_A)$  is the probability that $n_0$ is a root of an infinite $(k_1$-1)-ary 
subtree, $1-d_A$ is the probability that $n_0$ is an autonomous node, 
$d_A$ is the probability that $n_0$ depends on a node $n{'}$ in network $B$,
and $\Psi_B(f_B)$ is the probability that $n{'}$ is in the $k_2$-core of 
network $B$. Following similar arguments, we obtain the equation of $f_B$ in 
terms of $f_A$ and $f_B$,
\begin{equation}
f_B=\Phi_B(f_B)\left[ (1-d_B)+ d_Bp\Psi_A(f_A)\right]. \label{f_A_B_2}
\end{equation}
\begin{figure}
\includegraphics[width=0.45\textwidth,angle=0]{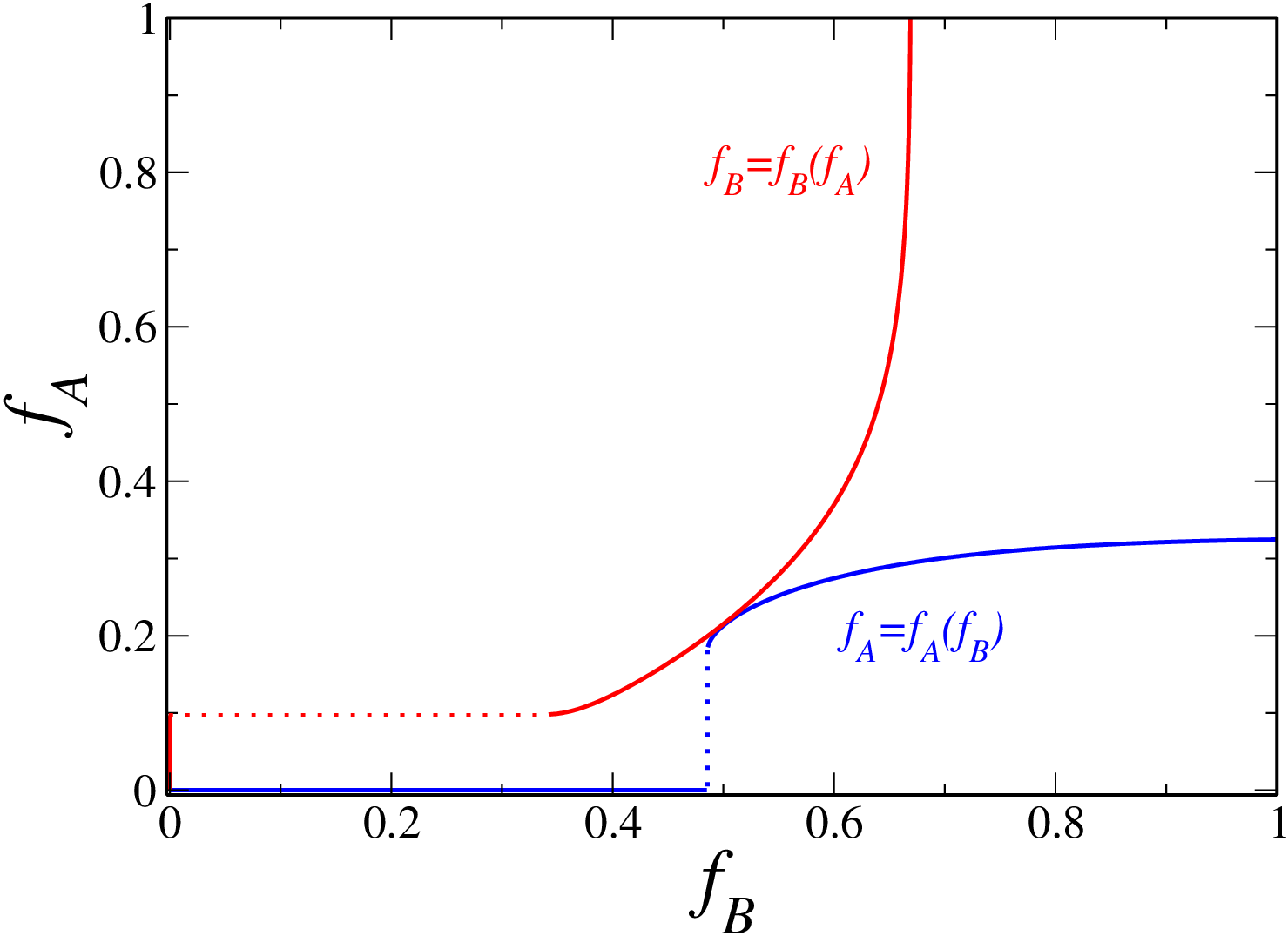}
\caption{\label{Fig8}(Color online) Graphical solution of Eqs.~(\ref{f_A_B_1}) 
and Eq.~(\ref{f_A_B_2}) for the $k$-core percolation with $\textbf{k}=(3,4)$ and $d=0.5$
 in two interdependent ER networks $A$ and $B$ with the average degree 10, where network
 $A$ is damaged initially under RA. 
The blue and red curves show, respectively, Eq.~(\ref{f_A_B_1}) and Eq.~(\ref{f_A_B_2})
 for the value of $p=p^{RA}_c(\textbf{k})$. The nontrivial solution of $f_A$ and $f_B$
appears at the critical value $p^{RA}_c(\textbf{k})=0.391$, at which the two curves intersect 
tangentially with each other, satisfying Eq.~(\ref{p_c}). When $p>p^{RA}_c(\textbf{k})$, these two curves will 
always have  a nonzero intersection and it serves as the  physical solution.  }
\end{figure}

Note that for any given value of $p$, $f_A$ and $f_B$ can be solved from 
Eqs.~(\ref{f_A_B_1}) and (\ref{f_A_B_2}) using Newton's method after 
choosing appropriate initial values.  We denote $M^A_{\textbf{k}}(p)$ and 
$M^B_{\textbf{k}}(p)$ as the probability that a randomly chosen node in 
network $A$ and $B$ belongs to the mutually connected $\textbf{k}$-core, 
respectively, and they satisfy
\begin{equation}
\begin{cases}
M^A_{\textbf{k}}(p)=p\Psi_A(f_A) \left[1-d_A+d_A\Psi_B(f_B)\right],\\ \label{giant_cores}
M^B_{\textbf{k}}(p)=\Psi_B(f_B) \left[1-d_B+d_Bp\Psi_A(f_A)\right].
\end{cases}
\end{equation}
Note that the mutually connected  $\textbf{k}$-core is made up 
of the $k_1$-core in network $A$ (with its normalized size 
denoted by $M^A_{\textbf{k}}(p)$) and the $k_2$-core in network $B$ 
(with its normalized size denoted by $M^B_{\textbf{k}}(p)$).

\begin{figure}
\includegraphics[width=0.45\textwidth, angle = 0]{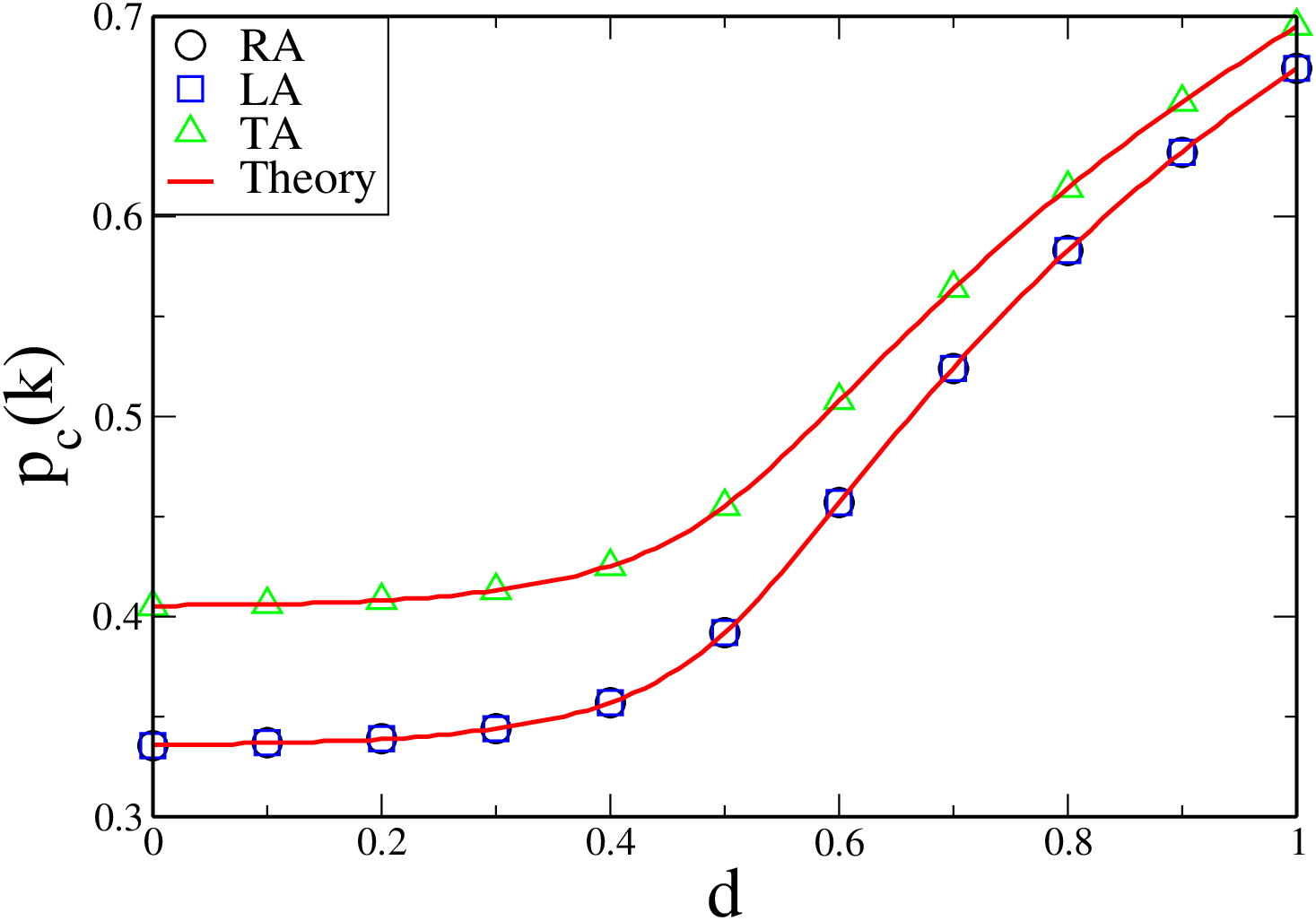}
\caption{(Color online)\label{Fig9} Percolation thresholds $p_c(\textbf{k})$ of two
 interdependent ER networks as a function of interdependence strength
  $d$ under RA, LA and TA with $\lambda=10$  and $\textbf{k}=(3,4)$.  
  Here solid lines are theoretical predictions and symbols ($\bigcirc$ for RA, 
 $\square$ for LA and $\bigtriangleup$ are for TA) are simulation results with network
  size of $N=10^6$ nodes. Note that for $d=0$ the results reduce 
  to the case of single networks with $k=3$, shown in Fig.~\ref{Fig4}.}
\end{figure}

The trivial solution $f_A=f_B=0$  for low occupation probability $p$ 
signifies the absence of a $\textbf{k}$-core in the system.  As $p$ 
increases, a nontrivial solution emerges in the critical case ($p=p^{RA}_c(\textbf{k})$)
 in which two curves $f_A=f_A(f_B)$ and $f_B=f_B(f_A)$ tangentially touch each other,
 i.e., 
\begin{equation}
\frac{df_A}{df_B}\cdot \frac{df_B}{df_A}=1 \label{p_c}
\end{equation}
which, together with Eqs.~(\ref{f_A_B_1}) and (\ref{f_A_B_2}),
gives the solution for $p^{RA}_c(\textbf{k})$ and the critical size of the 
mutually connected $\textbf{k}$-core. When $p>p^{RA}_c(\textbf{k})$, 
these two curves will always have  a nonzero intersection  that constitutes a
physical solution. For simplicity and without loss of generality, we use 
$d_A=d_B\equiv d$ throughout the rest of this paper.

(II) \textit{Localized Attack}: When LA is performed on the system of
 interdependent networks $A$ and $B$ described above, we find
an equivalent random network $E$ with a degree distribution $P(q{'})$ [from 
Eq.~(\ref{P(a)_LA})] such that after a random attack in which a fraction $1-p$ of
nodes in network $E$ are removed, the degree distribution of the remaining 
network is the same as the degree distribution of the remaining network resulting 
from an LA on network $A$.  Then by mapping the LA problem on interdependent 
networks $A$ and $B$ to a RA problem on a transformed pair of interdependent 
networks $E$ and $B$, we can apply the mechanism of RA on interdependent 
networks to solve $p_c^{LA}(\textbf{k})$ and the mutually connected 
$\textbf{k}$-core under LA.

(III) \textit{Targeted Attack}: Analogously, when TA is performed on the
interdependent networks $A$ and $B$ described above, we find
an equivalent random network $F$ with a degree distribution $P(q{'})$ [from 
Eq.~(\ref{P(a)_TA})] such that after a random attack in which a fraction $1-p$ of nodes in 
network $F$ are removed,  the degree distribution of the remaining network is 
the same as the degree distribution of the remaining network resulting 
from an TA on network $A$.  Thus, by mapping the TA problem on 
interdependent networks $A$ and $B$ to a RA problem on a transformed 
pair of interdependent networks $F$ and $B$, we can apply the mechanism of 
RA on interdependent networks to solve $p_c^{TA}(\textbf{k})$ and the mutually 
connected  $\textbf{k}$-core under TA in the case of $k$-core percolation.

\subsection{Results}

\subsubsection{Two interdependent  Erd\H{o}s-R\'{e}nyi networks}
\begin{figure}
 \includegraphics[width=0.45\textwidth, angle = 0]{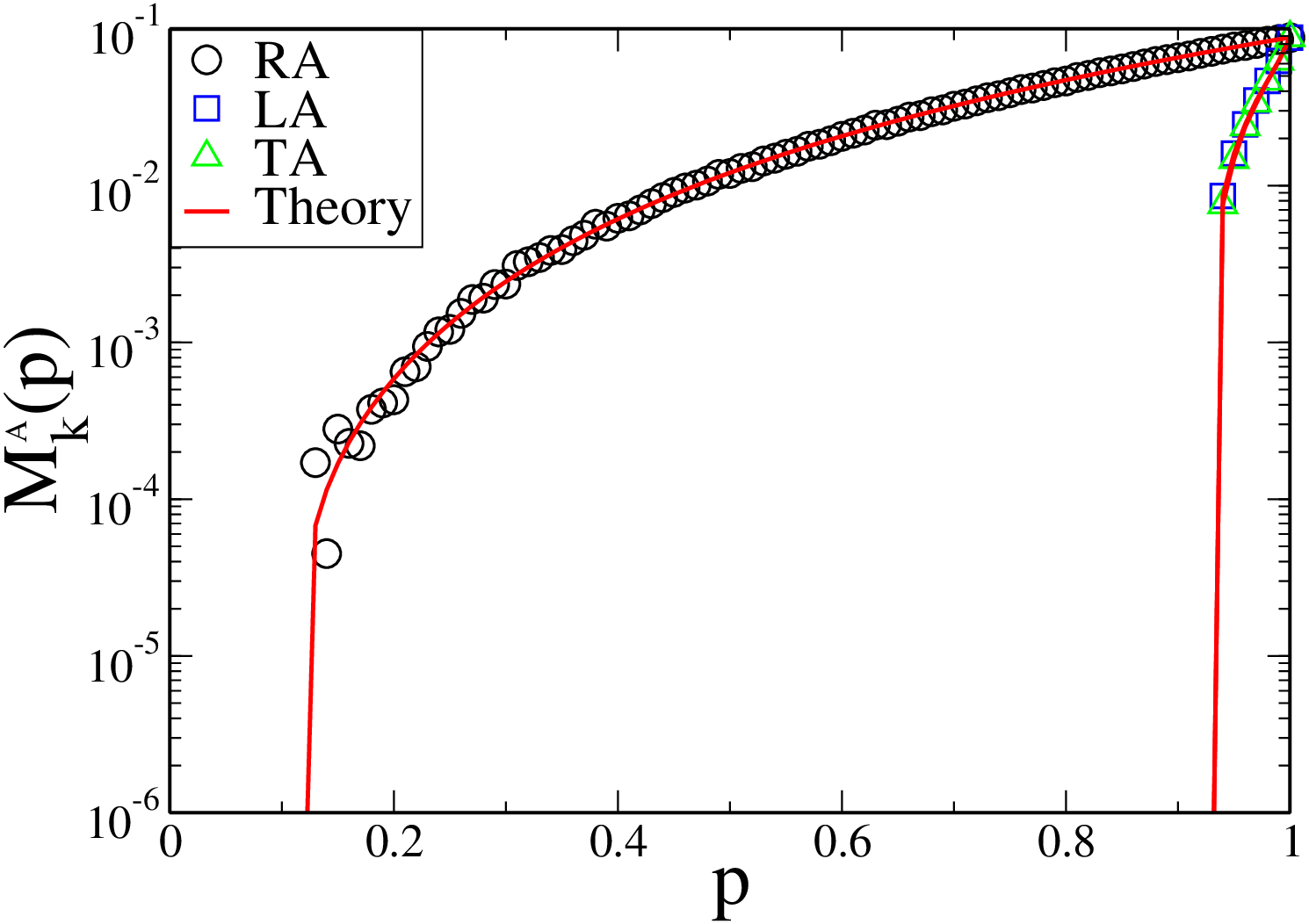}
\caption{\label{Fig10}(Color online)
  Sizes of the $k$-core of network $A$,
  $M_{\textbf{k}}^A(p)$, as a function of the fraction of unremoved nodes, $p$, for two
  partially interdependent SF networks with $d=0.5$,  $\gamma=2.3$, $q_{min}=2$,
  $q_{cut}(N)=1000$ and $\textbf{k}=(3,4)$.
  Here solid lines are theoretical predictions, from Eq.~(\ref{giant_cores}) for RA and
  its counterparts of LA and TA for $\alpha=1$, and symbols are simulation
  results with network size $N=10^6$, under RA ($\bigcirc$), LA ($\square$) 
  and TA ($\bigtriangleup$).}
\end{figure}
We start with two partially interdependent networks in which the degrees
both follow the same Poisson distribution and exert a RA on 
network $A$, initiating a $k$-core percolation pruning process that
continues until equilibrium is reached. We then follow the same procedure
with the same set-up but this time using a LA and TA to initiate the pruning
process.  Figure~\ref{Fig7} shows the $k$-core $M_{\textbf{k}}^A(p)$ 
of network $A$ as a function of the occupation probability $p$ under RA, LA and TA 
(with $\alpha=1$)  in the context of 
$k$-core percolation with $d=0.5$, $\textbf{k}=(3,4)$ and $\lambda=10$. 
The simulation results agree well with the theoretical results,
and there are first-order percolation transitions  in all attack
scenarios. As in single ER networks, note that $p^{RA}_c(\textbf{k} )$ is 
equal to $p^{LA}_c(\textbf{k})$ and 
both are smaller than $p^{TA}_c(\textbf{k})$. 

\begin{figure}
\includegraphics[width=0.45\textwidth,angle=0]{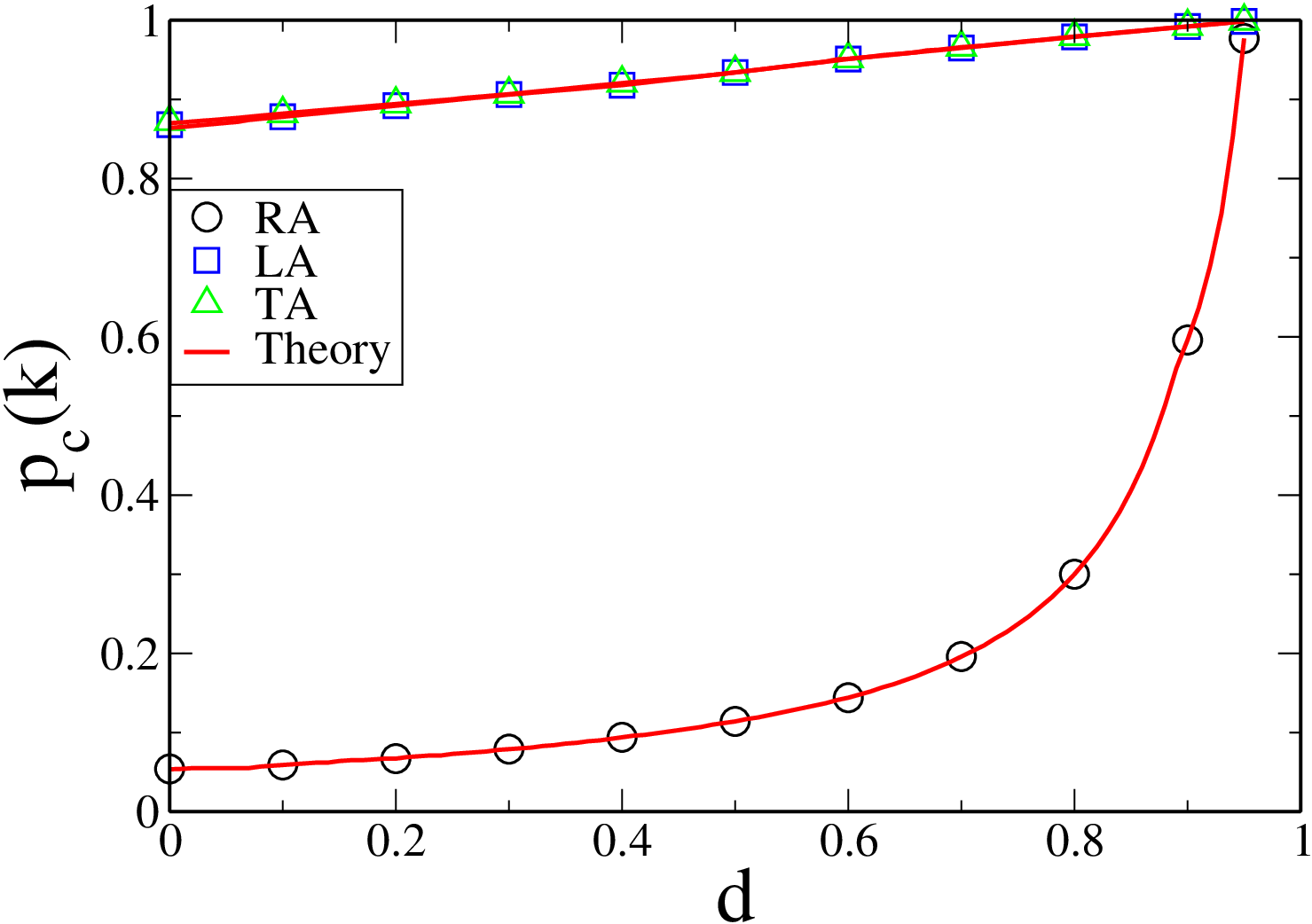}
\caption{\label{Fig11}(Color online) Percolation thresholds $p_c(\textbf{k})$ of two
  interdependent SF networks as a function of interdependence strength $d$ under RA, 
  LA and TA with $\alpha =1$, $q_{min}=2$, $q_{cut}(N)=1000$, $\gamma=2.3$
  and $\textbf{k}=(3,4)$.  
  Here solid lines are theoretical predictions and symbols ($\bigcirc$ for RA, 
 $\square$ for LA and $\bigtriangleup$ are for TA) are simulation results with network
  size of $N=10^6$ nodes. Note that for d=0 the results reduce to the case of 
  single networks with k=3, seen in Fig.~\ref{Fig6}.}
\end{figure}

Figure~\ref{Fig8} shows for instance the critical solution of Eqs.~(\ref{f_A_B_1}) and 
(\ref{f_A_B_2}) for the case of RA shown in Fig.~\ref{Fig7}.
When $p<p^{RA}_c(\textbf{k})$, the two curves representing Eqs.~(\ref{f_A_B_1}) and 
(\ref{f_A_B_2}) correspondingly intersect only at the origin,  and this always 
renders a zero-sized $k$-core $M_{\textbf{k}}^A(p)$ according to Eq.~(\ref{giant_cores}). 
A $k$-core $M_{\textbf{k}}^A(p)$ first arises discontinuously at 
$p=p^{RA}_c(\textbf{k})$, when these two curves tangentially touch each other at 
a nonzero intersection at $({f_A}_c, {f_B}_c)$, satisfying Eq.~(\ref{p_c}).  As $p$ 
increases further above $p^{RA}_c(\textbf{k})$,  $M_{\textbf{k}}^A(p)$
continues to exist because of the presence of a nonzero intersection that serves as the 
nontrivial solution of Eqs.~(\ref{f_A_B_1}) and (\ref{f_A_B_2}). Similar procedures 
are applied to the LA and TA scenarios as well and the corresponding $p^{LA}_c(\textbf{k})$ 
and $p^{TA}_c(\textbf{k})$ are obtained, respectively. 

Next we obtain the relationship between the robustness of the network system, i.e., the threshold
$p_c(\textbf{k})$, under 
three types of attacks and the interdependence strength $d$ in the context of $k$-core 
percolation. Figure~\ref{Fig9} shows how the percolation thresholds $p_c(\textbf{k})$ 
under RA, LA and TA (with  $\alpha =1$), change with $d$ where $\textbf{k}=(3,4)$ 
and $\lambda=10$ for two ER networks.  As seen in Fig.~\ref{Fig9}, when $d$ 
increases from 0  to 1,  $p^{RA}_c(\textbf{k})$, $p^{LA}_c(\textbf{k})$ and $p^{TA}_c(\textbf{k})$
increase accordingly, which means that
the higher the level of interdependence between networks $A$ and $B$,
the less resilient they are against attacks. Note that $d=0$ corresponds to the case in which there is no 
interdependence between networks $A$ and $B$ and the thresholds $p_c(\textbf{k})$
reduce to those shown in Fig.~\ref{Fig4} at $k=3$. For each $d$ value, $p^{RA}_c(\textbf{k})=p^{LA}_c(\textbf{k})<p^{TA}_c(\textbf{k})$,
which indicates that in the context of $k$-core percolation, RA and LA exert the same level of
damage to the structure of an ER network, but that TA produces more severe damage 
to an ER network. Similar results are reported in the context of ordinary percolation 
on interdependent ER networks \cite{yuan2015breadth,huang2011robustness}. 

\subsubsection{Two interdependent scale-free networks}
We construct two interdependent networks in which the degrees in
each follow the same power law distribution. Figure~\ref{Fig10} 
shows the $k$-core $M_{\textbf{k}}^A(p)$ of network $A$ as a 
function of the occupation probability $p$ under RA, LA and TA (with 
$\alpha=1$) under $k$-core percolation with $\textbf{k}=(3,4)$ and 
$\gamma=2.3$. The simulation results agree well with the theoretical 
results, and there is first-order percolation transition behavior in all attack
scenarios. Note that $p^{LA}_c(\textbf{k})$ is  approximately equal to 
$p^{TA}_c(\textbf{k})$ and they both are significantly larger than $p^{RA}_c(\textbf{k})$. 
As in single SF networks, the LA process can easily spread from the seed 
node to high degree hubs in few steps and therefore greatly disintegrates 
the core structure of the network, similar to the TA process.  
This is in strong contrast to the case of ER networks  in which  most
nodes have degrees close to the average degree and therefore for 
the RA and LA processes, nodes of high degrees are less likely to be removed compared to
the TA process.

Next we compare the robustness of the network 
system under each of the three types of attacks as a function of the interdependence strength 
$d$ in the context of $k$-core percolation. Figure~\ref{Fig11} shows how 
the percolation thresholds $p_c(\textbf{k})$ under RA, LA and TA (with $\alpha =1$), 
change with $d$ where $\textbf{k}=(3,4)$ and $\gamma=2.3$ 
for two SF networks.  Here in Fig.~\ref{Fig11}, as $d$ increases from 0  to 1, 
$p^{RA}_c(\textbf{k})$, $p^{LA}_c(\textbf{k})$  and $p^{TA}_c(\textbf{k})$ 
increase accordingly, which means that the more interdependent networks $A$ ad $B$ are on each other,
the less resilient they will be against attacks. Note that the $d=0$ case corresponds to the scenario shown in 
Fig.~\ref{Fig6} at $k=3$. For each $d$ value, $p^{LA}_c(\textbf{k})\approx p^{TA}_c(\textbf{k})>p^{RA}_c(\textbf{k})$,
which indicates that in the context of $k$-core percolation, LA and TA (with $\alpha=1$)
 exert approximately the same level of damage to the structure of a SF network whereas 
 RA produces less severe damage to a SF network. Similar results are 
reported in the context of ordinary percolation on SF networks \cite{huang2011robustness,shao2015percolation}.

\section{Conclusions}
 We have studied  and compared the robustness of 
both single and interdependent networks under three types of 
attacks in the context of $k$-core percolation.  We show that 
interdependence between networks makes the system more 
vulnerable than their single network counterparts. In addition,  
we map a network under LA and TA into an equivalent network 
under RA, solve analytically the $k$-core percolation problem,
and show how the initial attack type affects the robustness of 
networks. In general, TA exerts the most damage. In particular,  
LA and RA cause equal damage to ER networks whereas in ultrasmall 
networks like SF networks, LA causes much more damage 
than RA does. These findings hold for both single networks
and interdependent networks. 
\section*{Acknowledgments}

We wish to thank DTRA, NSF, the European MULTIPLEX,
ONR, and the Israel Science Foundation for financial support. Y.D. 
acknowledges support from the NSFC (Grant No. 71201132) and
the DFME (Grant No. 20120184120025).

\bibliographystyle{apsrev}

\bibliography{References}

\end{document}